\documentclass[useAMS,usenatbib]{mn2e}
\usepackage[dvips]{graphicx}
\usepackage[]{txfonts}
\def\simless{\mathbin{\lower 3pt\hbox
{$\rlap{\raise 5pt\hbox{$\char'074$}}\mathchar"7218$}}}   
\def\simmore{\mathbin{\lower 3pt\hbox
{$\rlap{\raise 5pt\hbox{$\char'076$}}\mathchar"7218$}}}   
\def\Msun{{\rm M}_\odot}                                       
\newcommand{\be}{\begin{equation}}
\newcommand{\ee}{\end{equation}}

\title{Signatures of a Maxwellian Component in Shock-Accelerated Electrons in GRBs}

\author[Dimitrios Giannios and Anatoly Spitkovsky]
{Dimitrios Giannios\thanks{E-mail: giannios@astro.princeton.edu
 (DG)} and Anatoly Spitkovsky\\
Department of Astrophysical Sciences, Peyton Hall, Princeton
  University, Princeton, NJ 08544, USA}

\begin{document}
\date{Received / Accepted}
\pagerange{\pageref{firstpage}--\pageref{lastpage}} \pubyear{2009}

\maketitle

\label{firstpage}

\begin{abstract}
Recent particle-in-cell simulations suggest that a large fraction
of the energy dissipated in a relativistic shock is deposited into a Maxwellian
distribution of electrons that is connected to the  
high-energy power-law tail. Here, we explore the
observational implications of such a mixed thermal-nonthermal particle
distribution for the afterglow and prompt emission of gamma-ray bursts.
When the Maxwellian component dominates the energy budget, the afterglow
lightcurves show a very steep decline phase followed 
by a more shallow decay when the characteristic synchrotron frequency crosses the observed
band. The steep decay appears in the X-rays at $\sim$100 sec after the burst
and is accompanied by a characteristic hard-soft-hard spectral evolution
that has been observed in a large number of early afterglows. If internal shocks
produce a similar mixed electron distribution, a bump
is expected at the synchrotron peak of the $\nu f_\nu$ spectrum. 
\end{abstract}


\section{Introduction} 
\label{intro}

After the prompt emission phase is over, the relativistic ejecta responsible
for the gamma-ray burst (GRB) drive a relativistic shock into the circumburst medium.
The afterglow emission that follows the GRB is believed to come from the
shocked external medium. Electrons are thought to be accelerated at the shock
front and radiate throughout the electromagnetic spectrum via  
synchrotron and synchrotron-self-Compton (SSC) mechanisms (see Piran 2005 for a review).

Particle acceleration in relativistic shocks is poorly understood. 
Most of the theoretical studies of GRB afterglows parameterize the distribution
of particles downstream of the shock by assuming that a fraction
$\epsilon_e$ of the energy dissipated at the shock goes to accelerating {\it all} 
electrons into a pure power-law distribution over a wide range of
energies (Paczynski \& Rhoads 1993; Sari et al. 1998).
The possibility that only a fraction $\zeta_e$ of the electrons are 
accelerated and the rest thermalize with temperature
$\Theta\equiv kT/m_ec^2 \sim \Gamma$, where $\Gamma$ is the bulk Lorentz 
factor of the shocked plasma, has also been considered (Eichler \& Waxman 2005).  

Recent particle-in-cell (PIC) simulations have begun to shed light on the acceleration
processes at work in relativistic shocks. The picture that emerges for the
particle distribution downstream from the shock is somewhat different than what has 
been traditionally assumed. Diffusive shock acceleration
does appear to operate in such shocks, accelerating a small fraction of particles (few $\%$)
into a power-law distribution that carries some $\sim 10\%$ of the 
dissipated energy. The rest of the electrons are, however,
found to be heated to temperature 
$\Theta\sim \Gamma m_p/m_e$ carrying most of the 
dissipated energy (Spitkovsky 2008a,b;  
Martins et al.\ 2009).

In this work, we explore the observational implications of such 
a mixed thermal-nonthermal particle distribution. We show how the
synchrotron emission spectrum and lightcurves are modified compared to conventional afterglow models. 
We suggest that the
spectral and temporal properties of the early X-ray lightcurves 
that show very steep decay may be produced with such a particle distribution.   
We also briefly discuss the possible implications of a mixed electron
distribution on the prompt GRB emission.

\section{Modeling the electron distribution downstream from the shock}

The GRB afterglow emission is believed to come from the shock driven by the ejecta  
into the circumburst medium. Since the afterglow is 
steadily observed over a wide range of frequencies ranging from the radio to
X-rays (Costa et al. 1997; Frail et al. 1997; van Paradijs et al. 1997;
and likely extends into the $\gamma$-rays), a broad particle distribution
downstream of the shock is needed to explain observations (Sari et al. 
1998; see also Paczynski \& Rhoads 1993; Waxman 1997; Wijers et al. 1997). 
A power-law distribution in energy for the particle number $N_e\propto \gamma^{-p}$ 
with index $p\sim 2-2.5$ is typically inferred (e.g., Waxman
1997; Galama et al. 1998). Particle acceleration through repeated shock
crossings is a promising mechanism to accelerate particles to such 
a distribution (e.g., Gallant, Achterberg \& Kirk 1999). 

Consider a relativistic shock which has a relative Lorentz factor  $\Gamma\gg 1$
between the shocked and unshocked fluid. 
A rather economical, and popular, parameterization of the particle 
distribution downstream of the shock assumes that all the electrons  
are accelerated into a power-law distribution 
for $\gamma>\gamma_{min}$. When $p>2$, most of the energy of the electrons
resides at the low end of the distribution. Then, $\gamma_{min}$ is found from
the total available energy per particle assuming that a fraction $\epsilon_e$
of the energy dissipated by the shock goes into the electrons, resulting in $\gamma_{min}\sim
\epsilon_e \Gamma m_p/m_e$. In this parameterization, the distribution has a sharp cutoff below
$\gamma_{min}$. 

Eichler \& Waxman (2005) considered the possibility of only a fraction 
of the electrons being accelerated to a nonthermal distribution. The rest were
assumed to isotropize and thermalize at a characteristic temperature $\Theta\sim
\Gamma$ given by the upstream kinetic energy of the {\it electrons} with no
extra heating from the ions. In the Eichler \& Waxman (2005) approach, the 
particle distribution is peaked at $\gamma\sim \Gamma$ and $\gamma\sim
\Gamma m_p/m_e$ with most of the energy carried by the nonthermal component.
The nonthermal electrons in this model are responsible for most of the emission
(with the possible exception of the early radio afterglow).  

Recent PIC simulations of relativistic shocks have revealed both the stochastically accelerated 
and Maxwellian components in the downstream electron distribution
(see Spitkovsky [2008b] and Sironi \& Spitkovsky [2009] for simulations of pair shocks, 
and Spitkovsky [2008a] and Martins et al.\ [2009] for electron-ion simulations).
Simulations of relativistic shocks in weakly magnetized electron-ion plasma
show that the thermalized electrons are substantially heated with respect to their upstream bulk flow energy,
receiving a large fraction of the 
dissipated energy from the ions. The electron thermal distribution is smoothly connected 
to the low end of the power-law component (see Fig.~1). The thermal component is a robust 
prediction of all simulations, and represents the downstream state of the bulk
flow. It can be depleted if the transfer of energy to the power-law component
is very efficient. Current simulations show that in shocks that accelerate
particles, the power law tail can take of the order of $10-20\%$ of the flow
energy. As the present simulations necessarily run for a limited amount of
time, it is possible that the steady state has not been achieved
yet. Therefore, it makes sense to explore the observational signatures
of the mixed thermal-nonthermal distribution for a range of acceleration efficiencies. 
 
\begin{figure}
\resizebox{\hsize}{!}{\includegraphics[angle=270]{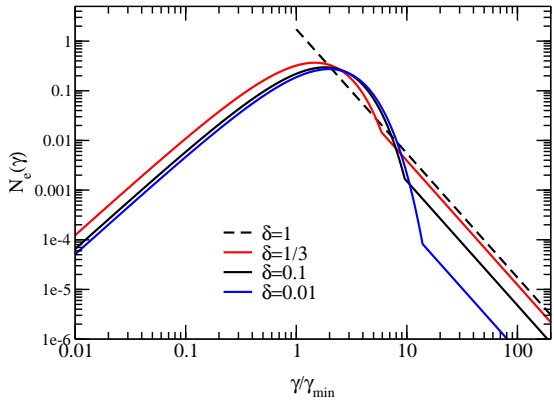}}
\caption[] {The electron-number distribution for different values
of the fraction $\delta$ of the total energy that goes into the
power-law component. For $\delta=1$, the pure power-law distribution
is reproduced (dashed line). The most recent PIC simulations indicate that
$\delta \sim 0.1$ (solid, black line). For $\delta\simless 1/3$, the
Maxwellian component in the electron distribution is pronounced.
\label{fig1}}
\end{figure}

In this work, we define the normalized electron distribution downstream of the 
shock by a continuous function connecting a relativistic Maxwellian 
to a power-law at $\gamma=\gamma_{nth}$:
\begin{eqnarray}
N_e(\gamma, \Theta)&=&CN_e^{th}(\gamma,\Theta),\quad {\rm for}\quad
\gamma\le\gamma_{nth}, \nonumber\\&&\label{N_e}\\
N_e(\gamma,
\Theta)&=&CN_e^{th}(\gamma_{nth},\Theta)(\gamma/\gamma_{nth})^{-p},\quad {\rm for}
 \quad  \gamma>\gamma_{nth}, \nonumber
\end{eqnarray}
where $N_e^{th}(\gamma,\Theta)=\gamma^2\exp(-\gamma/\Theta)/2\Theta^3$
is the Maxwell distribution in the $\Theta\gg 1$ limit (which is of interest
here) and $C$ is a normalization constant. The ratio $\gamma_{nth}/\Theta$
controls the fraction $\delta$ of the total energy that resides in the nonthermal component:
\be
\delta\equiv \frac{\int_{{\gamma_{nth}}}^{\infty}\gamma N_e(\gamma,\Theta)d\gamma}
{\int_1^{\infty}\gamma N_e(\gamma,\Theta)d\gamma}.
\label{delta}
\ee

Furthermore, we assume that a fraction $\epsilon_e$
of the energy dissipated in the shock is picked up by the electrons. 
The average (random) Lorentz factor per particle is
\be
<\gamma>\equiv
\int_1^{\infty}\gamma N_e(\gamma,\Theta)d\gamma=\epsilon_e\Gamma\frac{m_p}{m_e}.
\ee
Using the last expression with eqs.~(\ref{N_e}), (\ref{delta}) the electron
distribution is determined (i.e., $\Theta$ and $\gamma_{nth}$ can be calculated) 
after choosing $\delta$, $\epsilon_e$, $\Gamma$ and $p$.

The mixed thermal-nonthermal distribution reduces to the pure power-law
for $\delta=1$. In this case $\gamma_{min}=\gamma_{nth}=(p-2)\epsilon_e\Gamma m_p/(m_e
(p-1))$ which defines the minimum cutoff of the distribution. In the limit of
negligible nonthermal component $\delta\ll 1$, $\Theta=<\gamma>/3=\epsilon_e\Gamma
m_p/3m_e$.

\section{Synchrotron spectrum}

Plasma instabilities that lead to the formation of the
collisionless shock are also responsible for the 
amplification of magnetic fields in the shock transition
(Medvedev \& Loeb 1999). Whether these small scale fields
survive at large distances downstream from the shock is 
still an open question (Chang et al. 2008, Keshet et al. 2009, Gruzinov 2008;
{Medvedev \& Zakutnyaya 2009}). 
This is, of course, very important, since the
fields must remain strong on macroscopic scales for the 
synchrotron model to explain the afterglow
observations (Rossi \& Rees 2003, Waxman 2006). 

In this paper, we assume that the magnetic fields survive far downstream
from the shock and facilitate the synchrotron and SSC
radiation. In general, synchrotron dominates the X-ray and softer emission, while  
inverse Compton (IC) dominates in the $\gamma$-ray band 
(e.g., see Fan et al.\ 2008 and references therein).
Here, we will neglect the IC component and only focus on the new features
that appear in the synchrotron emission due to the thermal
component in the electron distribution.

We assume that a fraction $\epsilon_B\sim 10^{-2}$
of the energy dissipated at the shock remains in
the magnetic fields at macroscopic distances downstream from the shock. 
For a fluid moving with $\Gamma\gg 1$ just behind the
shock (in the frame of the central engine), the energy density of the shocked fluid is
$e=4\Gamma^2n_{ext}m_pc^2$, where $n_{ext}$ is the number
density of the external medium (e.g., Sari et al. 1998).
The comoving magnetic field strength is thus 
\be
B^2=8\pi\epsilon_Be=32\pi\epsilon_B\Gamma^2n_{ext}m_pc^2.
\label{Bfield}   
\ee
 
Using the standard expressions for synchrotron emission of 
ultrarelativistic electrons (Rybicki \& Lightman 1979),
we calculate the spectrum for various values of $\delta$.
In Fig.~2 we show a ``slow cooling'' case, where the cooling frequency
lies above the characteristic synchrotron frequency. A reverse situation,
corresponding to the ``fast-cooling''
case, is shown in Fig.~3. The cooling frequency is defined as the peak synchrotron 
frequency for electrons whose 
synchrotron cooling timescale equals the expansion timescale of the system.

\begin{figure}
\resizebox{\hsize}{!}{\includegraphics[angle=270]{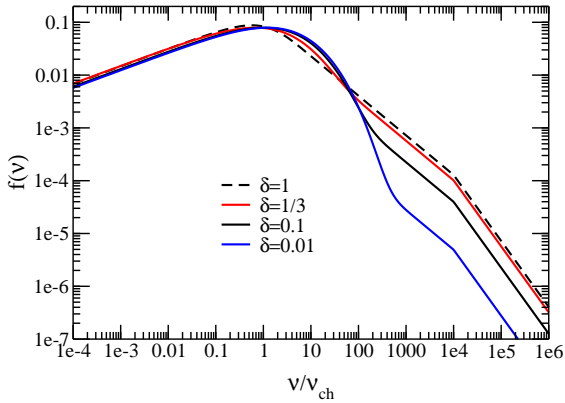}}
\caption[] {Synchrotron spectrum in the case of slow
cooling for different values of the fraction  $\delta$ of the energy
in the power-law component of the electron distribution. The cooling frequency is
taken to be $\nu_c=10^4\nu_{ch}$ ($\nu_{ch}$ is the characteristic 
synchrotron frequency). Self-absorption is not included in the calculation.
\label{fig2}}
\end{figure}
\begin{figure}
\resizebox{\hsize}{!}{\includegraphics[angle=270]{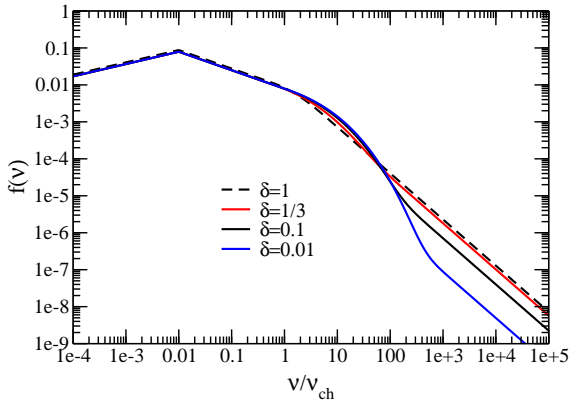}}
\caption[] {Synchrotron Spectrum in the case of fast
cooling for different values of the fraction of the energy
in the nonthermal component $\delta$. The cooling frequency is
taken to be $\nu_c=0.01\nu_{ch}$. The different particle distributions affect the spectral 
shape at the characteristic synchrotron frequency and the normalization
of the high-energy power-law emission.
\label{fig3}}
\end{figure}

The spectra for different $\delta$ are very similar below the characteristic frequency
$\nu_{ch}=\Gamma\gamma_{min}^2\nu_{cyc}$, where $\nu_{cyc}=eB/2\pi m_ec$ is 
the cyclotron frequency. The Maxwellian declines fast enough
for $\gamma\ll \Theta$ that the emission at lower frequencies comes
from particles close to the peak of the thermal distribution, which
is close to the low energy cutoff of the pure nonthermal power law for $\delta=1$.
Here we ignore the self absorption which appears in the
low-frequency part of the spectrum (typically, radio).
We estimate, however, that varying $\delta$ introduces a weak variation
in the self-absorption frequency. Thus, we do not discuss the self-absorption further.

The cooling frequency (or, equivalently, the Lorentz factor of electrons that
emit at the cooling frequency) does not depend on the shape of the particle distribution.
The difference in spectra due to particle distributions with different $\delta$
is clearly seen around the characteristic frequency
$\nu_{ch}$. For $\delta=1$ the characteristic
frequency appears as a break in the spectrum. For $\delta\ll 1$, above
the characteristic frequency (determined by the temperature of the thermal
component) there is a sharp decline of the emission that is followed by hardening at
higher energies. The sharp decline is not described by a power-law
spectrum. The hardening corresponds to the emission coming from the 
nonthermal component of the electron distribution.

This particular shape around the characteristic frequency is reflected in the
afterglow lightcurves. When the characteristic
frequency crosses the observer band, new afterglow features appear
as we discuss in the following section.

\section{Afterglow lightcurves}

We consider a blastwave that decelerates due to the accumulation of mass
from the circumburst medium. 
The bulk Lorentz factor of the blastwave depends on the total 
energy $E$, distance $R$ and the external medium density (and profile)
\be
\Gamma(R)\simeq \Big(\frac{E}{M_{ext}c^2}\Big)^{1/2},
\ee
where $M_{ext}=\int_0^R4\pi R'^2 n_{ext}m_pdR'$ is the total mass
accumulated from the external medium at distance R. 

We focus on two types of external medium: a constant density
medium, $n_{ext}=$const, and a wind-like profile $n_{ext}\propto 1/R^2$. 
Note, that we keep $E$ constant, i.e., we ignore the energy injection 
that appears to be needed to explain the shallow decay segments 
of the X-ray lightcurves (Zhang et al. 2006; Nousek et al. 2006). Such
inclusion is straightforward but does not affect our main points of
discussion.  

Although the modeling of the afterglow emission involves several parameters, most of them have
been extensively explored in the literature. Here we focus on
the new effects on the lightcurves from varying the fraction 
$\delta$ of the energy in the nonthermal component. 
We compute the lightcurves for $E=10^{53}$ erg, $\epsilon_e=0.3$, $\epsilon_B=0.01$, 
and $p=2.5$, which we will refer to as {\it reference values} of the parameters.
For constant density of the external medium we use $n_{ext}=1$
cm$^{-3}$ as a reference value. For the wind case, the
density profile is calculated assuming a spherical stelar wind of $\dot{M}=10^{-5}\Msun$/year
and velocity $v_w=10^8$ cm/sec expected from a Wolf-Rayet progenitor (see Li \&
Chevalier 1999). 

In computing the afterglow lightcurves, we use the standard approach
described in Sari et al. (1998) and assume a burst at luminosity distance
$d=10^{28}$ cm. In Figs.~4 and 5  we show the resulting lightcurves in the optical
and in Fig. 6 the lightcurve at 1 keV. 

The lightcurves are very similar for different
$\delta$ until the characteristic frequency crosses the observed
band. For $\delta=1$ the crossing appears as a single break in the lightcurve.
On the other hand, for $\delta\simless 0.1$ there is a 
steep decline of the lightcurve followed by a break that leads to 
a more shallow decline. During the steep decline the 
spectrum shows a characteristic curvature (see next section).
The time at which the steep decline
 is interrupted by a more shallow one occurs when the nonthermal  
electrons dominate the emission in the observed band. This
shape of the lightcurve is a robust expectation from a mixed  
thermal-nonthermal electron distribution for $\delta\simless 0.1$.

\begin{figure}
\resizebox{\hsize}{!}{\includegraphics[angle=270]{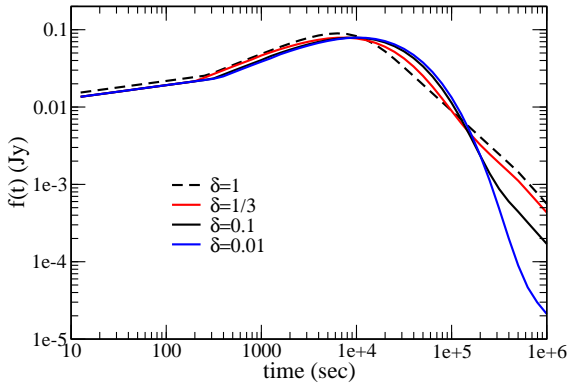}}
\caption[] {Synchrotron lightcurve in the optical band for the reference 
values of parameters and constant density external medium. 
The various curves correspond to different values of $\delta$.
The jet-spreading effects  (resulting in a break in the afterglow
lightcurve, so called jet break) are not included in the calculation.
  
\label{fig4}}
\end{figure}
\begin{figure}
\resizebox{\hsize}{!}{\includegraphics[angle=270]{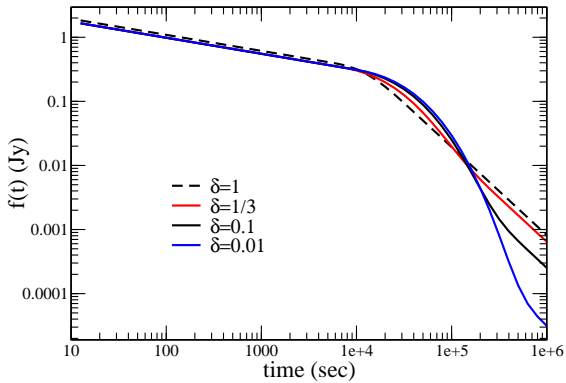}}
\caption[] {Same as Fig. 4 but for the wind-like external medium.
\label{fig5}}
\end{figure}
\begin{figure}
\resizebox{\hsize}{!}{\includegraphics[angle=270]{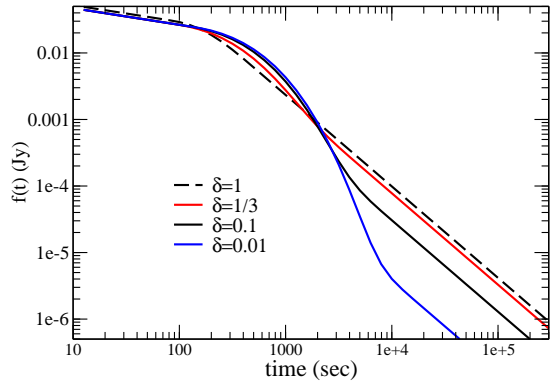}}
\caption[] {X-ray lightcurve for the reference values of parameters and
constant density external medium. A wind-like medium results in very similar
lightcurves (not shown for clarity of the plot). The jet-spreading effects
(resulting in the jet break) are not included in the calculation.
\label{fig6}}
\end{figure}

\subsection{Application to the early, steep X-ray decay}

Very steep decline in the early X-ray afterglow has been frequently
observed hundreds of seconds after the GRB (Nousek et al. 2006). 
The steep decays were initially attributed to off-axis GRB emission 
(high-latitude model; Kumar \& Panaitescu 2000). This interpretation predicts, however, a
specific relation between the temporal and spectral power-law indexes
during the very steep decay that is not in agreement with that observed in
the majority of the afterglows (O'Brien et al. 2006)\footnote{{ Zhang et
al. (2009) modified the high-latitude model by 
allowing for a non-power-law spectrum upon the cessation of 
the prompt emission phase. This model can account for the observed
spectral evolution during the steep decline of the X-ray afterglows
{\it if} at the end of the prompt emission the spectrum is steepening
just above the XRT band.}}. 

Zhang et al. (2007) did a systematic study of 44 steeply decaying
X-ray afterglows for which  time resolved spectra are available. They 
found that 11 X-ray tails did not show significant spectral evolution
with time and are compatible with the high-latitude emission model.
The rest show a clear spectral evolution not expected
from the high-latitude emission model. Out of those 33 afterglows,
16 have smooth tails (i.e., no flaring activity). The spectral index
of the smooth tails showed a characteristic steepening with time in the X-rays from
spectral index ($f_\nu\sim \nu^{-\beta}$) $\beta\sim 0.5$ to $\beta\sim 2$ 
(see Fig.~2 of Zhang et al. 2007). Moreover,
after the very steep decay is over, the spectrum hardens to $\beta \sim 1.2$.

All these temporal and spectral features appear naturally when the
characteristic frequency from a mixed distribution crosses the X-ray 
band for $\delta\sim 0.1$ and depend weakly on the 
parameters used to model the afterglow emission. 

Since $\nu_{ch}\sim\Gamma \gamma_{min}^2\nu_{cyc}$, the characteristic
frequency as a function of observer's time is (ignoring redshift corrections)
\be
\nu_{ch}\sim
10^6\epsilon_{e,0.3}^2\epsilon_{B,-2}^{1/2}E_{53}^{1/2}t_{obs}^{-3/2}\quad {\rm
eV}.
\ee
For deriving the last expression, we have used Eqs.~(4), (5) and that
the observer time{\footnote{ The expression for
$t_{\rm obs}$ is applicable within a factor of 2 for both wind and ISM
external media.}} $t_{obs}\sim R/4\Gamma^2c$.
Note that Eq.~(6) does not depend on the density of the external
medium and approximately applies for both constant density and wind-like 
media. The characteristic frequency crosses the
X-ray band marking the onset of the steep decay time $t_{sd}$ at
\be
t_{sd} \sim 300
\epsilon_{e,0.3}^{4/3}\epsilon_{B,-2}^{1/3}E_{53}^{1/3}\quad {\rm sec}
\ee 
after the burst (with the exact numerical value depending on $\delta$), 
leading to a steep decline of the lightcurve if $\delta \simless 0.1$
(see Fig.~6). 

During the steep decline there is a characteristic
softening of the spectrum, which hardens again after the the decay
enters the power-law phase (see Fig.~7). This spectral evolution
with time depends only on the fraction $\delta$ and not on the
external medium. The $\epsilon_{e}$, $\epsilon_{B}$ and $E$
parameters mainly affect {\it when} the steep decay (accompanied by the spectral
evolution) takes place but not the range over which the spectral index
varies.

{In Fig.~8 we overplot our model predictions with afterglow observations.
We select afterglows with good quality of data from the Zhang et
al. (2007) sample that show the typical hard-soft-hard evolution.}  The observed 
steepening of the spectrum from $\beta\sim 0.5$ (characteristic
of fast cooling) to $\beta\sim 2$ appears naturally for $\delta\sim 0.1$ while
the hardening to $\beta\sim 1$ after the steep decline is over
corresponds to the high-energy fast-cooling $p/2$ segment of the synchrotron 
spectrum\footnote{Note that one of the  models shown in Fig.~8 (see yellow
  line) shows an additional late-time raise of the spectral index from 
$\beta=0.55$ to $\beta=1.1$ because of the cooling break crossing the X-ray band.}.  

Energy injection in the blastwave may be needed to explain the shallow decay of the lightcurve
observed after the steep decay phase (Zhang et al. 2006; Nousek et
al. 2006). While the energy injection (that is not
included in our calculation) affects the temporal evolution of the lightcurves,
it does not affect the hard-soft-hard evolution of the spectrum that uniquely 
reflects properties of the particle distribution. We can, therefore, conclude
that in our interpretation of the steep decline, $\delta$ is constrained to
be $\sim 0.1$ in agreement with recent PIC simulations.

\begin{figure}
\resizebox{\hsize}{!}{\includegraphics[angle=270]{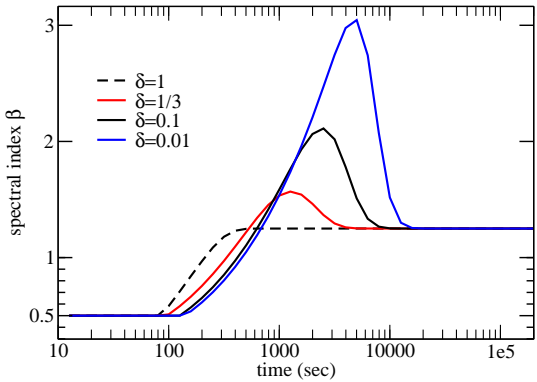}}
\caption[] {Evolution of the spectral index $\beta$ (where $f_\nu\sim
  \nu^{-\beta}$) of the $\sim$1 keV emission as a function of time for a 
constant density external medium and for different values of $\delta$.
A wind profile leads to very similar results. The
rest of the parameters are kept to their reference values.
The spectral softening takes place simultaneously
with the steep decay of the X-ray flux (see Fig.~6).\label{fig7}}
\end{figure}

\begin{figure}
\resizebox{\hsize}{!}{\includegraphics[angle=270]{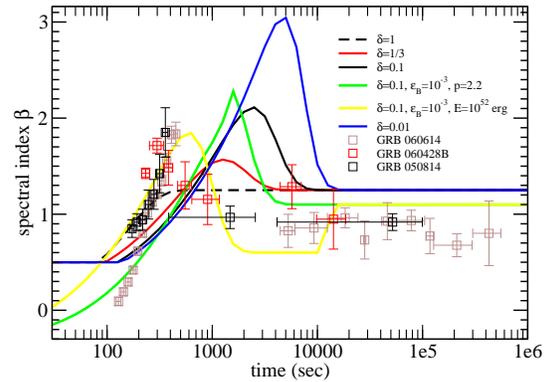}}
\caption[] {The evolution of the spectral index $\beta$ in the X-rays 
as a function of time for a constant density external medium as predicted by our model.
The dashed, red, black and blue lines are the same as in Fig.~7.
The green line shows the model predictions for $\delta=0.1$, 
$\epsilon_{B}=10^{-3}$, $p=2.2$ and the yellow line for $\delta=0.1$,
$E=10^{52}$ erg, $\epsilon_{B}=10^{-3}$, and $p=2.2$.  The
rest of the parameters are kept to their reference values.
Overplotted is the spectral evolution of the GRBs 050814, 060428B and
060614 observed with {\it XRT} and analyzed in Zhang et al. (2007).  
\label{fig8}}
\end{figure}

\section{Connections to the prompt GRB emission}

So far we have applied a mixed distribution of accelerated electrons
in relativistic shocks to the afterglow emission. 
figuThe dissipative process responsible for the prompt GRB emission
is more uncertain, with internal shocks (Paczynski \& Xu 1994; Rees \& Meszaros 1994) 
and magnetic dissipation (Usov 1992; Thompson 1994; Spruit et al. 2001;
Lyutikov \& Blandford 2003) being the leading candidates.
The prompt emission may be the result of synchrotron (Katz 1994), photospheric emission
({ Eichler \& Levinson 2000;}  M\'esz\'aros \& Rees 2000; Pe'er,  M{\'e}sz{\'a}ros \&
Rees 2006; Giannios 2006) or a combination of the photospheric (that dominates
the $\sim$1 MeV band) and synchrotron, SSC emission (contributing from the optical to the $\sim$
GeV band) because of gradual energy release (Giannios 2008).  

The internal shock model predicts mildly relativistic collisions. 
Furthermore, the colliding ejecta 
may be substantially magnetized. Mildly relativistic collisions of 
(potentially) magnetized plasma have not been studied in the same
detail by PIC simulations.
Keeping all these caveats in mind, we explore the possibility that the
prompt emission is produced due to synchrotron emission from internal shocks 
and that the distribution of the downstream particles is
a mixed, thermal-nonthermal one. {The effect to the prompt emission
coming from an injected strong Maxwellian component in the electron distribution 
has also been discussed in Baring \& Braby (2004) (see also Pe'er et al. (2006)).}    

As an example, we consider two shells ejected with a time difference 
$\delta t=0.1 \delta t_{-1}$ sec.
For the ratio of their bulk Lorentz factors $\Gamma_2/\Gamma_1$ of a few, 
the shells collide at distance $R_{IS}\sim \Gamma_{sh}^2 c \delta t/5$, where $\Gamma_{sh}$
is the Lorentz factor of shocked plasma (e.g., Bosnjak et al. 2009).
One can check that for $\epsilon_B=0.01$, $\epsilon_e=0.3$, $\Gamma_{sh}=100$,
the characteristic energy of the synchrotron emission is $E_{ch}\sim 40 $ keV

\begin{figure}
\resizebox{\hsize}{!}{\includegraphics[angle=270]{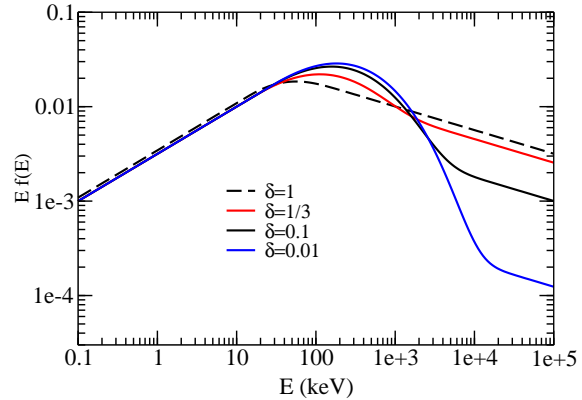}}
\caption[] {The fast-cooling synchrotron spectrum in $E f(E)$
representation (arbitrary units) for a shell collision with parameters
discussed in the text. A clear thermal-like bump appears 
when $\delta\simless 0.1$, followed by the high-energy emission with
spectral slope of $p/2$. { The low-frequency slope is that of 
fast-cooling electrons, i.e., with spectral slope $\beta=1/2$.}
\label{fig9}}
\end{figure}

Fig.~9 shows the fast-cooling synchrotron spectrum for different $\delta$
for the aforementioned internal shock parameters.
For the pure power-law electron distribution ($\delta=1$),
the emission spectrum appears as  smoothly connected power laws. 
A bump at the peak of the $Ef(E)$ spectrum is pronounced for 
$\delta\simless 0.1$ in addition to the low- and high-energy power laws. 
A similar feature has been
found by Baring \& Braby (2004) for a pronounced Maxwellian component
of the downstream electrons (although they adopted a different 
shape for the particle distribution). Note also that the 
peak of the spectrum increases by a factor of several for $\delta\simless 0.1$.

There is evidence  for ``thermal-like'' excess components 
in a number of bursts (Ryde 2005; { Ryde \& Pe'er 2009})\footnote {Note, however, that
the Ryde (2005) analysis fits the spectra with a blackbody and a single
power-law component while synchrotron emission from the mixed electron 
distribution results in a bump surrounded by
power-law high- and low-energy emission with {\it different} spectral
slopes.}.
Furthermore, a fraction of the bursts show a very steep decline
above the peak of the spectrum. These bursts might be connected
with very weak nonthermal acceleration in the shocks. The latter
is, for example, possible if the colliding ejecta are pre-magnetized
with large scale toroidal fields (Sironi \& Spitkovsky 2009).
On the other hand, a sample of bright bursts does not show any evidence 
for a significant Maxwellian component in the electron distribution
when their spectrum is fitted with a synchrotron model (Baring \& Braby 2004).   
The strength of the bump can be used to constrain
$\delta$ in the synchrotron-internal shock interpretation of the
GRB emission.  
 
{
The spectral slope bellow the peak of the synchrotron spectrum is that of fast
  cooling electrons, i.e. $\beta=1/2$. This slope is too soft in comparison to that
observed in the majority of the bursts. This is a well-known problem of the
synchrotron interpretation of the prompt GRB emission and is not ``cured'' by the
presence of a strong Maxwellian component in the electron distribution.
Synchrotron spectra with spectral slope $\beta\sim 0$ are expected bellow the
synchrotron peak provided that the inverse Compton cooling takes place
in the Klein Nishina regime (Derishev, Kocharovsky \& Kocharovsky 2001; Wang et al. 2009). Harder
spectra can be result of photospheric emission and/or heating of the
electrons over a timescale much longer than the cooling timescale
(slow cooling model; Ghisellini \& Celotti 1999; Stern \& Poutanen 2004; Pe'er
et al. 2006; Giannios \& Spruit 2007; Vurm \& Poutanen 2009).}

\section{Discussion/conclusions}     

Recent PIC simulations of relativistic shocks show that
the electron distribution forms a thermal component downstream of 
the shock that receives the largest fraction of the dissipated energy in
addition to the nonthermal, power-law component (Spitkovsky 2008a,b; Martins et al. 2009).
Here, we considered the effect of the thermal component on the
afterglow spectra and lightcurves and on the prompt GRB emission.

A strong Maxwellian component introduces new phenomenology when
the characteristic frequency of the synchrotron spectrum
crosses the observed band. Instead of a break in the lightcurve predicted  
by the pure power-law model for the electron distribution
(e.g., Sari et al. 1998), we find
a steep temporal decline followed by a break and a more shallow 
decay (see Figs.~4, 5, 6). During these phases the spectrum shows 
a characteristic hard-soft-hard evolution (Fig.~7). The steep decline
appears in the X-rays at $\sim$hundreds of seconds after the burst 
 {\it independently} of the external medium density and profile
(see Eq. (7)). 

The very steep decay observed in the early X-ray lightcurves
has spectral and temporal properties 
well studied thanks to the XRT on board of {\it Swift} (Nousek et al. 2006;
Zhang et al. 2007). In the majority of bursts, the steep decay shows a characteristic
``hard-soft-hard" spectral evolution that is in agreement to that expected 
from our model when the thermal component in the electron
distribution contains $\sim$10 times more energy than 
the nonthermal one (see Fig.~8; in agreement with recent PIC simulations).

Similar steep decline and spectral evolution is expected
when the characteristic frequency crosses other bands. 
From Figs.~4, 5 one can see that the steep decay appears at around 
1 day after the burst in the optical band (or slightly later if 
there is energy injection). During the steep decline, the spectrum
is predicted to deviate from a power-law. There are several optical afterglows
showing breaks and steep decline at $\sim 1$ day after the burst.
Unfortunately, the time coverage of the optical lightcurves is often sparse 
and there are other potential physical sources for breaks (end of the
energy injection, jet break) that complicate the interpretation
of the observations (Liang et al. 2008). It remains to be seen if 
optical observations support the mixed electron distribution that we propose.

Although the self-Compton emission is not included in our calculations,
we expect a similar steep decay signature when the Comptonised
component of the characteristic synchrotron frequency $\nu^{IC}_{ch}\sim
\gamma_{min}^2\nu_{ch}$ crosses the observed band. Depending on parameters, 
the steep decay can take place at $\sim$1000 sec in the $\sim$GeV band 
that is now accessible to observations thanks to the {\it FERMI} mission (see, e.g., 
Fan et al. 2008).  

Provided that the prompt GRB emission is the result of internal shocks,
similar considerations for the distribution of accelerated particles
can be applied to the prompt GRB itself.
If the large fraction of the energy goes into the Maxwellian component, 
the synchrotron peak of the prompt emission spectrum should exhibit 
a ``bump'' in addition to the typical fast-cooling synchrotron
spectrum from nonthermal (power-law) electrons that may be common in GRBs
(Ryde 2005), but not universal (Baring \& Braby 2004).

\section*{Acknowledgments}
We thank Bin-Bin Zhang for providing the link to observational data. 
DG acknowledges support from the Lyman Spitzer Jr. Fellowship awarded by the
Department of Astrophysical Sciences at Princeton University. AS acknowledges
support from the Alfred P. Sloan foundation fellowship and NSF grant  AST-080738.

\end{document}